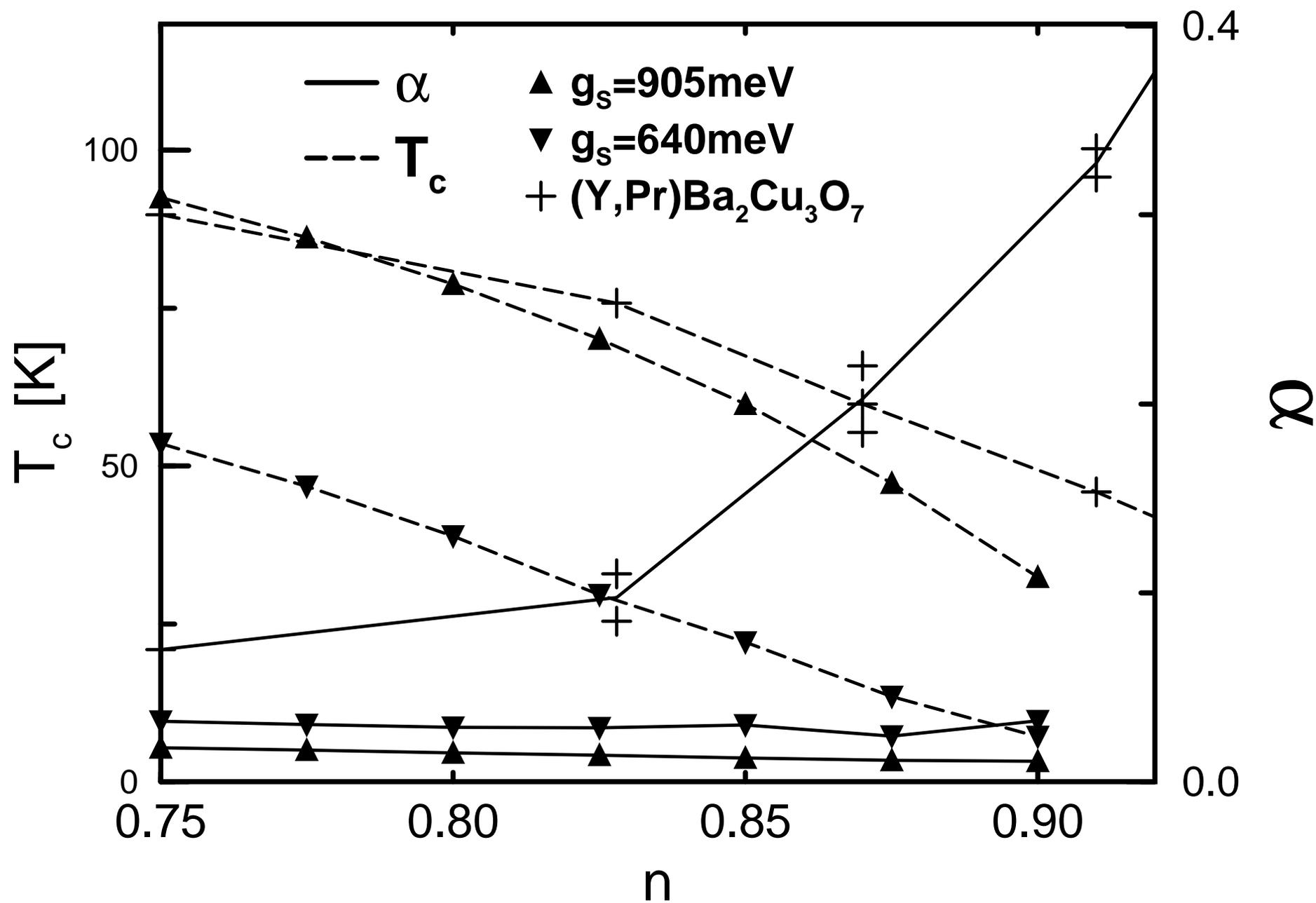

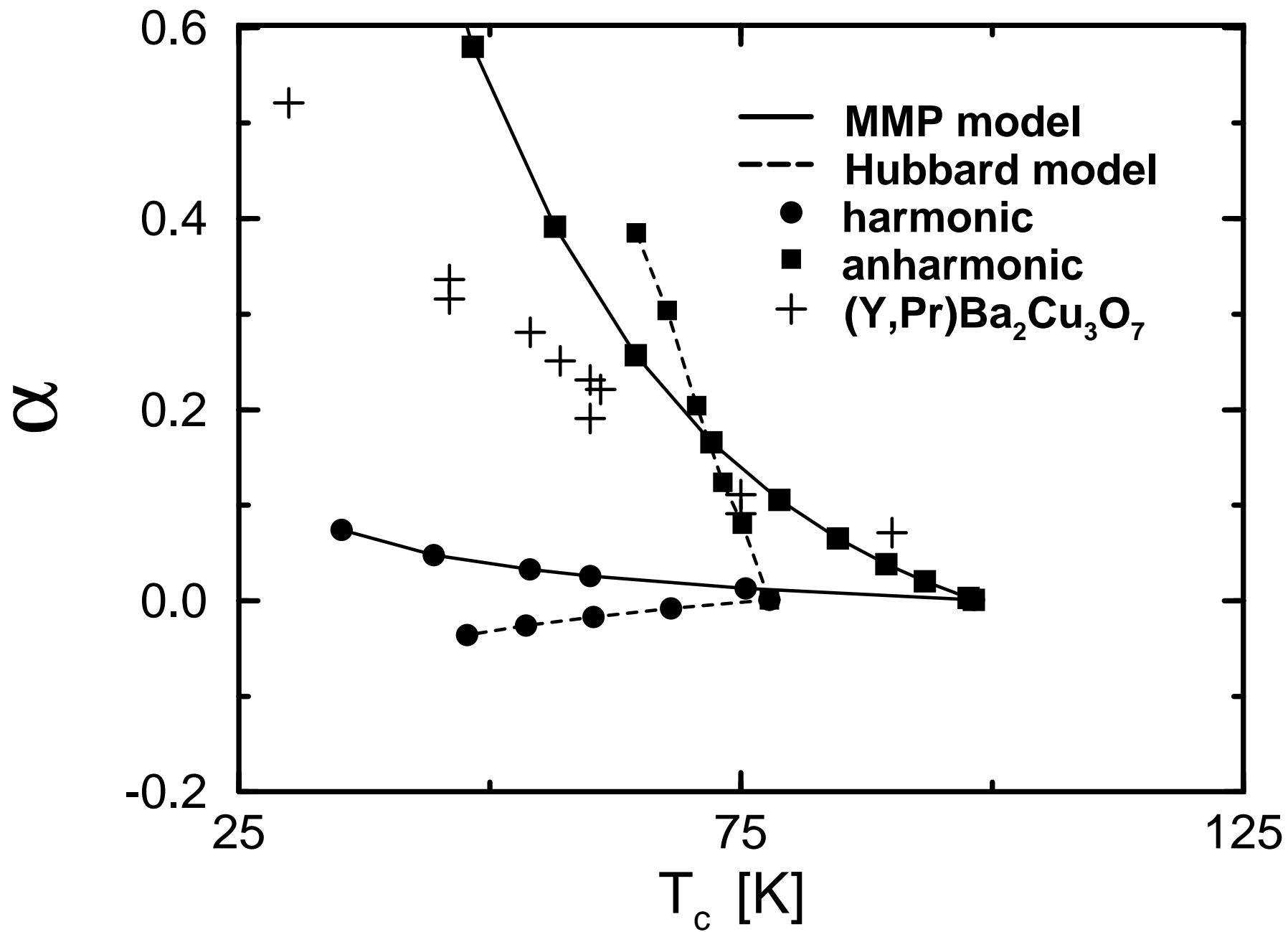

# THE ISOTOPE EFFECT
# IN d-WAVE SUPERCONDUCTORS

H.-B. Schüttler and C.-H. Pao

Center for Simulational Physics

Department of Physics and Astronomy

University of Georgia

Athens, Georgia 30602

Abstract

Based on recently proposed anti-ferromagnetic spin fluctuation exchange models for $d_{x^2-y^2}$-superconductors, we show that coupling to harmonic phonons *cannot* account for the observed isotope effect in the cuprate high-$T_c$ materials, whereas coupling to strongly anharmonic multiple-well lattice tunneling modes *can*. Our results thus point towards a strongly enhanced *effective* electron-phonon coupling and a possible break-down of Migdal-Eliashberg theory in the cuprates.

A growing, but not uncontroverted body of experimental evidence currently suggests that $YBa_2Cu_3O_7$ and, possibly, other cuprates are $d_{x^2-y^2}$-superconductors.[1,2] On the theoretical side, $d$-wave superconductivity would probably rule out *conventional* phonon-mediated pairing,[3] while supporting an electronic pairing mechanism, most prominently anti-ferromagnetic (AF) spin fluctuation exchange models.[2,4,5] Yet, except for certain "optimal" doping concentrations, many cuprates, including $YBa_2Cu_3O_7$, exhibit a quite noticeable isotope effect,[6] the classical hallmark of the superconducting electrons' coupling to the lattice vibrational degrees of freedom.

Starting from recent diagrammatic AF spin fluctuation models,[2,4,5] we will show here that the observed *order of magnitude* of the isotope exponent[6] $\alpha \sim 1$ does in fact imply an extraordinarily large electron-phonon (EP) coupling parameter $\lambda \gg 1$. For coupling to harmonic phonon systems, such coupling strengths would be well in excess of upper limits imposed by structural stability arguments and, independently, by the observed $T_c$ and normal-state transport scattering rates. However, the required $\lambda$'s *are* consistent with a coupling to local, large-amplitude lattice tunneling excitations, arising in a strongly anharmonic multiple-well lattice potential. Barring experimental problems in the isotope measurements,[6] our results thus point towards a very strong enhancement of the *effective* EP coupling, strong enough, in fact, to cause a break-down of the very approximation upon which the conventional Migdal diagrammatic theory is based. As such, our results point towards the possibility that lattice vibrational degrees of freedom play a central role in the low-energy electronic properties of the cuprates.

We start from the linearized Migdal-Eliashberg equations[3] for the self-energy $\Sigma$ and pair wavefunction $\Phi$ at wavevector $k$ and Matsubara frequency $i\nu = (2m+1)\pi i T$ (in energy units) at the transition temperature $T = T_c$,

$$\begin{aligned}\Sigma(k', i\nu') &= -\frac{T}{N} \sum_{k,i\nu} V_-(k'-k, i\nu'-i\nu) G(k, i\nu) \\ \Phi(k', i\nu') &= -\frac{T}{N} \sum_{k,i\nu} V_+(k'-k, i\nu'-i\nu) |G(k, i\nu)|^2 \Phi(k, i\nu).\end{aligned} \quad (1)$$

Here, $k \equiv (k_x, k_y)$ is summed over an $N \equiv L \times L$ grid covering the Brillouin zone of a two-dimensional (2D) square lattice. The single-electron Green's function $G$ is obtained self-consistently from $G(k, i\nu) = [i\nu - \epsilon_k - \Sigma(k, i\nu)]^{-1}$, assuming a single, 2D electron band $\epsilon_k = -2t_1(cos(k_x) + cos(k_y)) - 4t_2 cos(k_x) cos(k_y) - \mu$ with chemical potential $\mu$ and 1st and 2nd neighbor hopping $t_1$ (=250meV)[5a] and $t_2$, respectively. For singlet $d$-wave pairing, the effective electron-electron interaction potentials in Eqs.(1) are $V_\pm(q, i\omega) = \pm g_s^2 \chi_s(q, i\omega) + V_p(q, i\omega)$. This includes a spin-fluctuation term with coupling constant $g_s^2$ and dynamical spin susceptibility $\chi_s(q, i\omega)$,[2,4,5] and a phonon term of the general form $V_p(q, i\omega) = -U_p f(q) \Omega_q^2 / (\omega^2 + \Omega_q^2)$, assuming a single phonon branch with an energy dispersion $\Omega_q$ and a form factor $f(q)$, normalized so that $U_p = -N^{-1} \sum_q V_p(q, 0)$. The ratio $\bar{\lambda} \equiv U_p/B$ provides then a rough estimate for the dimensionless Eliashberg parameter $\lambda$,[3] averaged over the electronic bandwidth $B = 8t_1$. We emphasize that the isotopic mass- ($M$-) dependence of $V_p$ enters *only* via $\Omega_q$, but *not* via $U_p$.[3]

Global structural stability of the undoped, $\frac{1}{2}$-filled-band cuprate parent compounds requires the phonon-mediated on-site attraction $U_p$ to be less than the Coulombic on-site repulsion, of order of the on-site Hubbard $U$-parameter.[7] If $U_p > U$, the conduction electrons would form local pairs (bi-polarons), rather than local magnetic moments. At $\frac{1}{2}$-filling, such local pairs would undergo ordering in a charge superlattice,[8] rather than forming the AF spin superlattice observed[1,2] in the undoped cuprates. This transition from local moment AF spin density order to local pair charge density order has been established in recent studies of Holstein-Hubbard and related EP models.[8]

Local structural stability in the doped systems, i.e. stability of the dopant induced carriers against polaronic self-localization,[8] requires an even more stringent upper limit $U_p < U_p^{(loc)} < U$. If $U_p > U_p^{(loc)}$, polaron formation would destroy (or, at the very least, strongly renormalize and broaden) the delocalized quasi-particle states whose existence is essential for the Migdal-Eliashberg approach. Holstein-Hubbard estimates[8a] give $U_p^{(loc)} \sim 2 - 3 \times t_1$ for a Hubbard $U \sim 8 - 12 \times t_1$ in the cuprates,[7] or, equivalently, $\bar{\lambda}^{(loc)} \equiv U_p^{(loc)}/B \sim 0.25-0.4$ and $\bar{\lambda}^{(glob)} \equiv U/B \sim 1-1.5$.

Our results for the $d_{x^2-y^2}$ superconducting transition temperature $T_c$ and its isotope exponent $\alpha \equiv -d\log(T_c)/d\log(M)$ are illustrated by Figs. 1 and 2 for the case of an Einstein phonon model with $\Omega_q \equiv \Omega_0 = const$, local coupling $f(q) \equiv 1$, and assumed isotopic mass dependence $\Omega_0 \propto M^{-1/2}$. Such a model could roughly represent, for example, the local coupling(s) to high-energy optical modes which are dominated primarily by the lightest atomic species, i.e. oxygen, in the cuprates with $\Omega_0 \lesssim 100\text{meV}$.[9] The $\chi_s$ in the MMP model[5], used in Figs. 1 and 2, is given by

$$\text{Im}\chi_s(q, \omega + i0^+) = \text{Im}\left[\chi_Q / \left(1 + \xi^2 |q - Q|^2 - i\omega/\omega_s\right)\right] \theta(\omega_c - |\omega|) \quad (2)$$

with $Q = (\pi, \pi)$ for $q_x, q_y \geq 0$. *Except for values explicitly stated* in Figs. 1 and 2, we have used $U_p = 1\text{eV}$, $\Omega_0 = 50\text{meV}$ and the band and $T$-independent spin fluctuation parameters from Table II of Ref. [5a], hereafter referred to as MP-II, for a hole doping concentration $x \equiv 1 - n = 25\%$. Eqs. (1) are solved accurately, without further approximations, by fast Fourier transform techniques.[10]

Fig. 1 shows $T_c$ and $\alpha$ as functions of the electron concentration $n$. To model roughly the doping dependence of $\chi_s$, $\omega_s(x)$ and $\xi(x)$ are varied with $x \equiv 1 - n$, like[5a] $\omega_s \propto \xi^{-2} \propto x$, and attain their MP-II-values at $x = 25\%$. Fig. 2 shows $\alpha$ vs. $T_c$ for the same model at fixed $x = 25\%$, with $T_c$ varied by increasing $U_p$ from 0 to 2eV. Fig. 2 also shows results for coupling to anharmonic phonon and a Hubbard-based spin fluctuation model, discussed below. The figures clearly demonstrate the *very general result* that coupling to harmonic phonons will substantially suppress the $d$-wave $T_c$ without, however, causing a large enough isotope effect to account for the *order of magnitude* of the experimental results, illustrated here by the $(Y, Pr)Ba_2Cu_3O_7$ data.[6a]

Note here that the MP-II spin fluctuation parameters, with $g_s = 640\text{meV}$, were carefully tuned[5a] to reproduce the observed NMR data, resistivity *and* $T_c \cong 90\ °\text{K}$ in $YBa_2Cu_3O_7$, *excluding* the effect of phonons. If we now *include* phonons, with a strength of only $U_p \cong 1\text{eV}$ (i.e. $\bar{\lambda} \cong 0.5$), say, we need to roughly double $g_s^2$ (see

Fig. 1) in order to maintain a $T_c \cong 90$ °K. Thus, we will also double, roughly, the normal-state model resistivity from spin fluctuation scattering (not even counting phonon scattering !), which already seriously compromises the agreement with the experimental resistivity data. So, in the MMP model,[5] Eq. (2), the observed *maximum* $T_c \cong 90$ °K, combined with the resistivity data for $YBa_2Cu_3O_7$, will again limit $U_p$ to $U_p \lesssim 1$eV.

We have generated overwhelming numerical evidence that the foregoing results are generic and robust, (i) by systematically varying, over wide ranges, all spin fluctuation, electron-phonon and band parameters (ii) by using different types of form factors $f(q)$, different types of and/or multiple $q$-dependent phonon branches (including e.g. 3D acoustic phonons), (iii) by modifying the AF spin fluctuation model (introducing e.g. spin gap effects into the MMP model or using other proposed types of spectral models[2,4ab,11]), and (iv) by adding impurity scattering.[11,12] While such modifications of the model may substantially affect $T_c$, they do not increase the overall magnitude of $\alpha$ significantly. We thus conclude that the observed magnitudes of $T_c$ *and* $\alpha$ in the cuprates cannot possibly be explained by a conventional Migdal-Eliashberg treatment of the conduction electrons' coupling to AF spin fluctuations and harmonic phonons.[13]

To understand why harmonic phonon exchange $V_p$ causes a *large* $T_c$ suppression, but only a *small* isotope effect, consider first its effect on the self-energy $\Sigma$ in a simple McMillan-type analysis,[3,4] generalized to the case of $d$-wave pairing. In the real-frequency domain, $i\nu \to \nu + i0^+$, $V_p$ contributes, via $V_-$, to the quasi-particle mass enhancement $Z = 1 - \partial_\nu \text{Re}\Sigma|_{\nu=0} \equiv 1 + Z_s + Z_p$ and to the quasi-particle damping $\text{Im}\Sigma \equiv \text{Im}\Sigma_s + \text{Im}\Sigma_p$, with "$s$" and "$p$" denoting the spin fluctuation and phonon contribution, respectively. Now, $Z_p \cong \lambda \sim U_p/B$ is independent of isotopic mass and, in McMillan theory, enters into $T_c$ roughly as[3,4] $T_c \sim \langle \omega \rangle \exp(-Z/\lambda_d)$ with a bare $d$-wave pairing strength[3,4] $\lambda_d$ and an appropriate average energy $\langle \omega \rangle$ of the $V_+$-spectrum.[3,4] Thus, $Z_p$ causes a large *isotope-independent* reduction in $T_c$, by roughly a factor $\exp(-\lambda/\lambda_d)$, which is quite consistent with our numerical

$-5-$

results for the $U_p$-dependence of $T_c$. On the other hand, the phonon contribution to the quasi-particle damping,[3] $|\text{Im}\Sigma_p| \sim \lambda \max_q(\Omega_q)$, is comparable to the phonon energy scale $\Omega_q$ and thus much smaller than the typically electronic magnitude of $\Sigma_s$. Therefore, $\text{Im}\Sigma_p$ causes an additional $T_c$-suppression, by pair-breaking, which *does* depend on isotopic mass, but is much smaller than the "intrinsic" $T_c$-suppression, due to $\Sigma_s$, thus causing only a small $\alpha$.[13] Treating the phonons solely in terms of an Abrikosov-Gorkov pair-breaking theory[14] (*without* inclusion of the mass enhancement effect !) is seriously flawed, since it severely underestimates the $T_c$-suppression caused by the phonons, for given magnitude of the isotope exponent. In this regard, the effect of phonons is quite different from that of impurities.[11,12]

In the case of a $q$-dependent $V_p(q,i\omega)$, $V_p$ may also contribute directly, via $V_+$, to the $d$-wave pairing potential. In McMillan theory,[3,4] the dominant effect is again an *isotope-independent* phonon contribution $\lambda_{d,p}$ to the bare $d$-wave pairing strength $\lambda_d \equiv \lambda_{d,s} + \lambda_{d,p}$ entering into $T_c \sim \langle\omega\rangle \exp(-Z/\lambda_d)$. The *isotope-dependent* contribution $\langle\omega\rangle_p$ to $\langle\omega\rangle \equiv \langle\omega\rangle_s + \langle\omega\rangle_p$ is again of order of the phonon energy scale, i.e. small compared to the spin fluctuation contribution $\langle\omega\rangle_s$, thus, again, giving only a small contribution to $\alpha$. A substantially larger $\alpha$ and smaller $T_c$-suppression *may* be obtained in the case of *phonon-mediated* $d$-wave pairing, i.e. if $V_p(q,i\omega)$ is attractive in the $d_{x^2-y^2}$-pairing channel,[12] with large enough $\lambda_{d,p} > 0$. Because of the expected cancellation effects in the presence of multiple phonon branches, this scenario seems unlikely, but should nevertheless be explored by first-principles $\lambda_{d,p}$-calculations.

Phonon renormalizations of the spin fluctuation exchange potential $g_s^2 \chi_s$ do also not provide a viable mechanism for obtaining a significantly larger $|\alpha|$, as illustrated by the Hubbard model results in Fig. 2. Here, both $\Sigma$ and $g_s^2 \chi_s$, as well as a less important charge fluctuation term $g_c^2 \chi_c$ in $V_\pm$, Eq. (1), were calculated self-consistently, from $G(k,i\nu)$, in the fluctuation exchange approximation to the 2D Hubbard model,[2,4a] using numerical renormalization group techniques.[4b] The

– 6 –

phonon term $V_p$ in Eq. (1) thus explicitly modifies $\chi_s$ and $\chi_c$, via $\Sigma$. The result, $|\alpha| \ll 1$, is explained by noting that, again, the dominant effect of $V_p$ on $\chi_s$ and $\chi_c$ comes from the *isotope-independent* mass renormalization $Z_p$.

Fig. 2 also shows results obtained for coupling to an anharmonic phonon system, consisting of independent, local anharmonic oscillators with one lattice displacement degree of freedom $u_j$ per lattice site $j$. Each $u_j$ is subject to a local Hamiltonian $h_j \equiv -(\hbar^2/2M)\partial^2/\partial u_j^2 + w(u_j)$ with a double-well potential $w(u_j) = 4\Delta_B[2(u_j/d)^4 - (u_j/d)^2]$. Here, $d$ is the distance between the two local $w$-minima. The chosen tunneling barrier height $\Delta_B = 26.89$meV and re-scaled atomic mass $\bar{M} \equiv Md^2/\hbar^2 = 0.3555$meV$^{-1}$ give a double-well tunneling splitting $\Omega_t \equiv E_1 - E_0 \cong 7.25$meV and a quasi-harmonic (single-well "phonon") excitation energy $\Omega_h \equiv E_2 - E_0 \cong 35.5$meV where $E_\ell$ denotes the $\ell$-th excited state energy of $h_j$. The $q$-independent $V_p$ was calculated as[15] $V_p(i\omega) = -C^2 \int_0^\beta d\tau \langle u_j(\tau)u_j(0)\rangle \exp(i\omega\tau)$, with the local correlation function $\langle u_j(\tau)u_j(0)\rangle$ obtained from numerical solutions for the eigenstates of $h_j$. $T_c$ in Fig. 2 was varied by raising the re-scaled coupling $\bar{C} \equiv Cd$ from 0 to[8a,15a] of 200meV and 300meV for the MMP[5] [Eq.(2)] and Hubbard[4] model, respectively. In contrast to harmonic phonon models, $\bar{\lambda} \equiv -V_p(0)/B \sim \bar{C}^2/(B\Omega_t)$, at $T=0$,[8a,15a] is strongly enhanced here and strongly isotope-dependent, since the tunneling splitting $\Omega_t$ varies exponentially with the isotopic mass $M$. For given $T_c$, this model thus leads to a much larger $\alpha$ and, for given $\alpha$, requires a much smaller bare coupling $C$ than the harmonic phonon models discussed above. These larger values of $\alpha$ can be achieved without substantially increasing the model's DC resistivity, since the phonon contribution to the quasi-particle damping, Im$\Sigma$, is, again, small compared to the spin fluctuation contribution. When studied as a function of doping concentration $x$, with $\xi$ and/or $\omega_s$ varied with $x$ as in Fig. 2, the isotope exponent in the anharmonic model typically increases in magnitude with decreasing $T_c$ as one moves away from the optimal doping. This behavior appears to be generic to the model and is in qualitative agreement with the data in the cuprates.

Substantial experimental evidence for a strongly anharmonic multiple-well lattice dynamics in the cuprates exists.[9] Polaron formation is a possible mechanism for generating anharmonic tunneling modes of sufficient effective coupling strength to the conduction electron system within physically reasonable parameter limits.[8] The foregoing treatment extends the Migdal-Eliashberg approach to model the two essential features of such a polaronic system, namely, multiple-well lattice anharmonicity and isotope-dependent electron mass renormalization.[8,15a] However, as polaronic instabilities may well cause a break-down of the Migdal approximation, the present treatment should be re-examined in a more general strong-coupling framework.[8]

In summary, based on a McMillan analysis and large body of numerical evidence, we conclude that harmonic phonon exchange causes a large $T_c$-suppression, but only a small isotope exponent $\alpha$, in $d$-wave pairing instabilities driven by AF spin-fluctuation exchange. Physical constraints on the EP coupling strengths limit $|\alpha|$ to theoretical values below 0.1, which is an order of magnitude smaller than observed values in non-optimally doped, reduced-$T_c$ cuprate materials. By contrast, acceptable values of $\alpha$ at reasonably large $T_c$'s *can* be obtained, with physically reasonable coupling constants, by exchange of strongly anharmonic lattice tunneling excitations. Such anharmonic lattice modes lead to a very strongly enhanced, isotope-dependent effective $\lambda$-parameter. As such, our results, combined with the experimental isotope data,[6] can be regarded as evidence for a very strongly enhanced effective EP coupling in the cuprates which may well cause a break-down of the Migdal approximation. The further development of strong-coupling approaches,[8] beyond Migdal-Eliashberg theory, will be crucial for gaining a deeper understanding of the role of phonons in the low-energy physics of the cuprate high-$T_c$ systems. Experimentally, it would be of great interest to study the isotope-dependence of other physical properties, for example the specific heat, in order to test whether the cuprates indeed exhibit a strong isotope-dependence of their electronic quasi-particle mass, as predicted by polaronic and anharmonic lattice models.

We would like to acknowledge helpful discussions with J. P. Franck, J. Martindale, P. Monthoux, D. Pines and D.J. Scalapino. This work was supported by the National Science Foundation under Grant Nos. DMR-8913878 and DMR-9215123 and by the University of Georgia Office of the Vice President for Research. Computing support from UCNS, University of Georgia, and NCSA, University of Illinois, is gratefully acknowledged.

FIGURE CAPTIONS

Fig.1. $d_{x^2-y^2}$ transition temperature $T_c$ and isotope exponent $\alpha$ vs. electron concentration $n$ for coupling to Einstein phonons in the MMP model, Eq.(2), with $\omega_s \propto \xi^{-2} \propto (1-n)$. Also shown are the $T_c$ and oxygen isotope data for $(Y, Pr)Ba_2Cu_3O_7$, from Ref.[6a], with $n = 0.75$ assumed at the maximum $T_c$.

Fig.2. Isotope exponent $\alpha$ vs. $d_{x^2-y^2}$ transition temperature $T_c$ in the Hubbard model, Ref.[4a,b], with $U/t_1 = 6$, $t_1 = 250$meV, $t_2 = 0$, and $n \cong 0.86$, and in the MMP model, Eq.(2), with MP-II parameters, for coupling to *harmonic* Einstein phonons, and for coupling to *anharmonic* tunneling excitations, with EP parameters given in text. Also shown are the oxygen-isotope-vs.-$T_c$ data for $(Y, Pr)Ba_2Cu_3O_7$, from Ref.[6a].